\documentclass[11pt,a4wide]{article}
\usepackage{newlfont}
\usepackage{amsmath}
\usepackage{graphics}
\usepackage{float}
\usepackage{graphicx}
\usepackage{epsfig}
\usepackage{multirow}
\usepackage{multicol}
\usepackage{verbatim}
\usepackage{pdflscape}
\usepackage{cite}
\usepackage{array}

\begin{document}

\title{Spectrum and Decay Properties of Bottomonium Mesons}
\author{Ishrat Asghar\thanks{e mail: ishrat.asghar@ue.edu.pk},~Nosheen Akbar\thanks{e mail: nosheenakbar@cuilahore.edu.pk} \\
\textit{$\ast$Department of Physics, University of Education Lahore, Faisalabad Campus, Faisalabad.}\\
\textit{$\dag$Department of Physics, COMSATS University Islamabad, Lahore Campus,}\\
{Lahore(54000), Pakistan.}}
\date{}
\maketitle

\section*{Abstract}
We calculate the spectrum and wave functions (WFs) of various states of bottomonium mesons ($b\overline{b}$) using a non-relativistic quark potential model (NRQPM). The calculated WFs are used to compute the radiative widths of various states of $b\overline{b}$. The strong decays widths of bottomonium states are also calculated using $^3P_0$ model by choosing simple harmonic oscillator wave functions (SHOWFs). The $\beta$ of SHOWFs for various states of the mesons are measured by fitting the numerical wave functions. The radiative and strong decay widths are used to calculate the branching ratios of $b\overline{b}$ mesons. We also compare our calculated masses and widths with available experimental data.

\section{Introduction}

Upsilon($\Upsilon$), a state of bottomonium meson, was observed first time in E288 experiment at Fermilab \cite{Herb} in 1977. The next newly discovered state of $b\overline{b}$ was the $3 P$ state that was observed in Large Hadron Collider (LHC) in 2011\cite{ATLAS12, ATLAS14}. Uptill now eighteen states of $b\overline{b}$ mesons have been observed in experiments at BaBar, Belle, CDF, D0, ATLAS, CMS and LHCb with lowest state mass equal to $9.3909 \pm 0.0028$ GeV and highest state mass equal to $11.019\pm0.008$ GeV. For theoretical investigation of this experimentally obtained data and to predict new states of bottomonium mesons, different approaches have been used.

Non-relativistic quark model \cite{Vijande05} is used to calculate the masses and decays of bottomonium mesons in refs. \cite{Segovia16, Segovia08}.  Martin-like potential model is used in ref.\cite{Shah} to calculate the masses and leptonic widths of $b\overline{b}$ and $c\overline{c}$ mesons. Relativistic quark potential model \cite{Godfrey85, GodfreyD3185, Godfrey86, Godfrey04, Godfrey05} is useed in refs. \cite{Godfrey15,wang18} to calculate the masses and decay properties of bottomonium mesons. Constituent quark model with the incorporation of spin dependant interaction is used in ref. \cite{zheng23} to calculate the masses and leptonic widths of various states of $b\overline{b}$ and $c\overline{c}$ mesons.

In this paper, we study the masses, radiative transitions, strong decays and branching ratios of $b\overline{b}$ meson upto higher states with $n L= 5 S, 4 P, 4 D, 1 F$. For this, we used non-relativistic quark potential model in the columbic plus linear form alongwith the incorporation of spin-spin and spin-angular momentum interactions to find the masses and WFs of $b\overline{b}$ mesons. Parameters are found by fitting the experimentally available masses of bottomonium, bottom and bottom-strange mesons with the model calculated masses by taking different values for coupling constant for each sector. The calculated WFs are used to calculate the E1 and M1 radiative widths. Strong decay widths are calculated with simple harmonic oscillator wave function (SHOWF) using $^3 P_0 $ model for ground and excited states of $b\overline{b}$ mesons. SHOWF depends on the parameter $\beta$. In Ref. \cite{ferretti18}, strong decays for open charm and open bottom flavour mesons are calculated by taking same value of $\beta$ for different flavoured mesons, but in the present paper, strong decay widths of all angularly excited $b\overline{b}$ states are calculated using different values of $\beta$ for different flavoured states.
Authors of ref.\cite{Godfrey15} used different values of parameter $\beta$ for different states of bottomonium mesons in the calculation of decay properties. They found $\beta$ by fitting the RMS radii of SHOWF to the corresponding WF of relativitic quark potential model. But, we find $\beta$ by fitting SHOWF with the numerically calculated WFs of non-relativistic potential model.
We combine radiative and strong widths to predict the branching ratios of all possible decay channels of $b\overline{b}$ states.

The paper is organized as follows. In section 2, the potential model is described which is used to calculate the mass and WF of different states of $b\overline{b}$ mesons. In Sec. 3, the expressions used for E1 and M1 radiative transitions are defined. The methodology for the calculation of the strong decay amplitudes using $^3P_0$ decay model is explained in Section 4. Results are discussed in Section 5; while the concluding remarks are given in Section 6.

\section{Potential model for bottomonium, charmed bottom and bottom mesons}

Following non-relativistic quark anti-quark potential model\cite{Godfrey05} is used to find the mass spectrum and WFs of $b\overline{b}$, strange-bottom and bottom mesons.

\begin{eqnarray}
V_{q\bar{q}}(r) &=&\frac{-4\alpha _{s}}{3r}+br+\frac{32\pi \alpha _{s}}{
9 m_{q} m_{\bar{q}}}(\frac{\sigma }{\sqrt{\pi }})^{3}e^{-\sigma ^{2}r^{2}}%
\mathbf{S}_{q}.\mathbf{S}_{\bar{q}}  \notag \\
&&+\frac{1}{m_{q} m_{\bar{q}}}[(\frac{2\alpha _{s}}{r^{3}}-\frac{b}{2r})%
\mathbf{L}.\mathbf{S}+\frac{4\alpha _{s}}{r^{3}}T].
\end{eqnarray}

$m_{q}$, $m_{\bar{q}}$ are the constituent masses of quark and anti-quark respectively. $\alpha _{s}$ is the strong coupling constant,$b$ is the string tension. Columbic interactions, spin-orbit interactions at short distance, and tensor
interactions are the result of one gluon exchange process; while spin-orbit interactions at large distances are the result of Lorentz scalar confinement.
The spin-spin $\mathbf{S}_{b}.\mathbf{S}_{\bar{b}}$, spin-orbit $\mathbf{L}.\mathbf{S}$, and tensor operators in $ \left\vert J,L,S\right\rangle $ basis are given by
\begin{eqnarray}
T& =\Bigg \{
\begin{array}{c}
-\frac{1}{6(2L+3)},J=L+1 \\
+\frac{1}{6},J=L \\
-\frac{L+1}{6(2L-1)},J=L-1.
\end{array}%
\end{eqnarray}

\noindent The values of parameters $\alpha _{s}$, $b$, $\sigma $, $m_{q}$, $m_{\bar{q}}$ are found by fitting the mass spectrum of bottomonium, strange-bottom and bottom mesons to the available experimental data of masses. This available data consists of eighteen states of bottomonium mesons, four states of strange-bottom meson and four states of bottom mesons given in Table 1 and Table 3. The best
fit values of these parameters are $b=0.1139\text{ GeV}^{2}$, $\sigma =0.6$ GeV, $ m_{b}=4.825$ GeV, $ m_{s}=0.41$ GeV, $ m_{u}= m_{d}=0.365$ GeV, $\alpha_{s}(b \overline{b})= 0.3339$, $\alpha _{s}(\mathbf{B}_s) = 0.738$, and $\alpha _{s}(\mathbf{B})= 0.92$.  To calculate the spectrum of various states of $b\overline{b}$ system we numerically solved the radial Schr$\ddot{\textrm{o}}$dinger equation given by
\begin{equation}
U^{\prime \prime }(r)+2\mu (E-V(r)-\frac{L(L+1)}{2\mu r^{2}})U(r)=0, \label{de}
\end{equation}

\noindent $\mu $ is the reduce mass of meson. Non-trivial solutions of the above equation, existing only for certain discrete values of energy ($E$), are found by the shooting method. Mass of a $b \overline{b}$ state is found by following expression:
\begin{equation}
m_{b\bar{b}}=2m_{b}+E,
\end{equation}

\begin{table}[H]
\centering
\renewcommand{\arraystretch}{0.6}
\caption{Masses of ground and excited states of bottomonium mesons. The SHO $\beta$ values are listed in the last column which are obtained by fitting SHO wave functions to the quark model wavefunctions.} \label{bottomonium masses}
\begin{tabular}{c c c c c c c}
\hline\hline
\hspace{0.3cm} nL \hspace{0.3cm} & \hspace{0.3cm} Meson \hspace{0.3cm} & \hspace{0.2cm} Our calculated mass \hspace{0.2cm} & \hspace{0.3cm} Expt. mass~\cite{PDG-22}& \hspace{0.3cm} $\beta$ \hspace{0.3cm}\\
\hspace{0.3cm}  \hspace{0.3cm} & \hspace{0.3cm}  \hspace{0.3cm} & \hspace{0.2cm} (GeV) \hspace{0.2cm} & \hspace{0.3cm} (GeV)& \hspace{0.3cm} (GeV) \hspace{0.3cm}\\
   \hline
1S & $\eta_b(1^1S_0)$    &9.5467   & $9.3987\pm 0.002$   &0.999 \\
   & $\Upsilon(1^3S_1)$  &9.5508   & $9.4603\pm 0.00026$  &0.996 \\
   \hline
2S & $\eta_b(2^1S_0)$    &9.9766   & $9.99\pm0.0035^{+0.0028}_{-0.0019}$    &0.756 \\
   & $\Upsilon(2^3S_1)$  &9.9778   & $10.023\pm 0.00031$  &0.754 \\
\hline
3S & $\eta_b(3^1S_0)$    &10.2315  &              &0.634 \\
   & $\Upsilon(3^3S_1)$  &10.2323  & $10.3552\pm 0.0005$ &0.633 \\
\hline
4S & $\eta_b(4^1S_0)$    &10.4365  &              &0.57 \\
   & $\Upsilon(4^3S_1)$  &10.437   & $10.5794\pm 0.0012$ &0.57 \\
\hline
5S & $\eta_b(5^1S_0)$    &10.6069  &              &0.53 \\
   & $\Upsilon(5^3S_1)$  &10.6073  & $10.8852^{+0.0026}_{-0.0016}$  &0.53 \\
\hline
1P & $h_b(1^1P_1)$       &9.905    & $9.8993\pm 0.0008$  &0.678\\
   & $\chi_{b0}(1^3P_0)$ &9.8919   & $9.85944\pm 0.00042\pm 0.00031$ &0.675 \\
   & $\chi_{b1}(1^3P_1)$ &9.9039   & $9.89278\pm 0.00026\pm 0.00031$ &0.675 \\
   & $\chi_{b2}(1^3P_2)$ &9.9108   & $9.91221\pm 0.00026\pm 0.00031$ &0.675 \\
\hline
2P & $h_b(2^1P_1)$       &10.1672  &$10.2598\pm 0.0012$  &0.611\\
   & $\chi_{b0}(2^3P_0)$ &10.1573  &$10.2325\pm 0.0004\pm 0.0005$  &0.61 \\
   & $\chi_{b1}(2^3P_1)$ &10.166   &$10.25546\pm 0.00022\pm 0.0005$ &0.61 \\
   & $\chi_{b2}(2^3P_2)$ &10.1711  &$10.26865\pm 0.00022\pm 0.0005$ &0.61 \\
\hline
3P & $h_b(3^1P_1)$       &10.3771  &              &0.558\\
   & $\chi_{b0}(3^3P_0)$ &10.3685  &              &0.558 \\
   & $\chi_{b1}(3^3P_1)$ &10.3754  &$10.5134\pm 0.0007$  &0.558 \\
   & $\chi_{b2}(3^3P_2)$ &10.3803  &$10.524\pm 0.0008$   &0.558 \\
\hline
4P & $h_b(4^1P_1)$       &10.5561  &              &0.523\\
   & $\chi_{b0}(4^3P_0)$ &10.5489  &              &0.522 \\
   & $\chi_{b1}(4^3P_1)$ &10.5551  &              &0.522 \\
   & $\chi_{b2}(4^3P_2)$ &10.5588  &              &0.522 \\
\hline
1D & $\eta_{b2}(1^1D_2)$ &10.0889  &              &0.589       \\
   & $\Upsilon_1(1^3D_1)$&10.0861  &              &0.588         \\
   & $\Upsilon_2(1^3D_2)$&10.0892  &$10.1637\pm 0.0014$  &0.588         \\
   & $\Upsilon_3(1^3D_3)$&10.0911  &              &0.588         \\
\hline
2D & $\eta_{b2}(2^1D_2)$ &10.3059  &              &0.555        \\
   & $\Upsilon_1(2^3D_1)$&10.3033  &              &0.554         \\
   & $\Upsilon_2(2^3D_2)$&10.3061  &              &0.554         \\
   & $\Upsilon_3(2^3D_3)$&10.3078  &              &0.554         \\
\hline
3D & $\eta_{b2}(3^1D_2)$ &10.4928  &              &0.522       \\
   & $\Upsilon_1(3^3D_1)$&10.4904  &              &0.521         \\
   & $\Upsilon_2(3^3D_2)$&10.493   &              &0.521         \\
   & $\Upsilon_3(3^3D_3)$&10.4946  &              &0.521         \\
\hline
4D & $\eta_{b2}(4^1D_2)$ &10.6497  &              &0.497        \\
   & $\Upsilon_1(4^3D_1)$&10.6477  &              &0.497         \\
   & $\Upsilon_2(4^3D_2)$&10.6498  &              &0.497         \\
   & $\Upsilon_3(4^3D_3)$&10.6511  &              &0.497         \\
\hline
1F & $h_{b3}(1^1F_3)$    &10.2305  &              &0.545        \\
   & $\chi_{b2}(1^3F_2)$ &10.2299  &              &0.544        \\
   & $\chi_{b3}(1^3F_3)$ &10.2309  &              &0.544         \\
   & $\chi_{b4}(1^3F_4)$ &10.231   &              &0.544         \\
\hline
2F & $h_{b3}(2^1F_3)$    &10.4239  &              &0.522        \\
   & $\chi_{b2}(2^3F_2)$ &10.4232  &              &0.522         \\
   & $\chi_{b3}(2^3F_3)$ &10.4242  &              &0.522         \\
   & $\chi_{b4}(2^3F_4)$ &10.4245  &              &0.522         \\
\hline
\hline
\end{tabular}
\end{table}

\section{Radiative transitions}
Radiative transitions are important to investigate the higher states of $b\overline{b}$ mesons. $E1$ radiative transitions from a $b\overline{b}$ meson to other $b\overline{b}$ meson state are calculated by using the following expression defined in ref.~\cite{Godfrey05}.
\begin{equation}
\Gamma_{E1}(n^{2S+1}L_J\rightarrow n'^{2S'+1}L'_{J'}+\gamma)=\frac{4}{3}C_{fi}\delta_{S S'}e_b^2 \alpha\mid < \Psi_f \mid r \mid \Psi_i>\mid^2 E_\gamma^3 \frac{E^{(b\overline{b})}_f}{M^{(b \overline{b})}_i}.  \label{E1}
\end{equation}
Here
 $E_\gamma$, $E^{b \overline{b}}_f$, and $M_i$ stand for final photon energy ($E_\gamma = \frac{M_i^2 - M_f^2}{2 M_i}$), energy of the final $b\bar{b}$ meson, and mass of initial state of $b\overline{b}$ meson respectively, and
 \begin{equation}
C_{fi}=max(L, L')(2 J'+1)\left \{
                           \begin{array}{ccc}
                             L' & J' & S \\
                             J & L & 1 \\
                           \end{array}
                         \right \}^2.
\end{equation}
$M1$ radiative transitions for a $b\overline{b}$ meson state to other $b\overline{b}$ meson state are calculated by the following expression~\cite{Godfrey05}:
\begin{equation}
\Gamma_{M1}(n^{2S+1}L_J\rightarrow n'^{2S'+1}L'_{J'}+\gamma)=\frac{4}{3}\frac{2J'+1}{2 L+1}\delta_{L L'}\delta_{S S'\pm 1}e_b^2 \frac{\alpha}{m^2_b}\mid < \Psi_f \mid \Psi_i>\mid^2 E_\gamma^3\frac{E^{(b \overline{b})}_f}{M^{(b \overline{b})}_i}. \label{M1}
\end{equation}

\section{Open Flavor Strong Decays}
\label{sect:open-flavor-strong-decays}
\begin{figure}
  \centering
  \includegraphics[width=10cm]{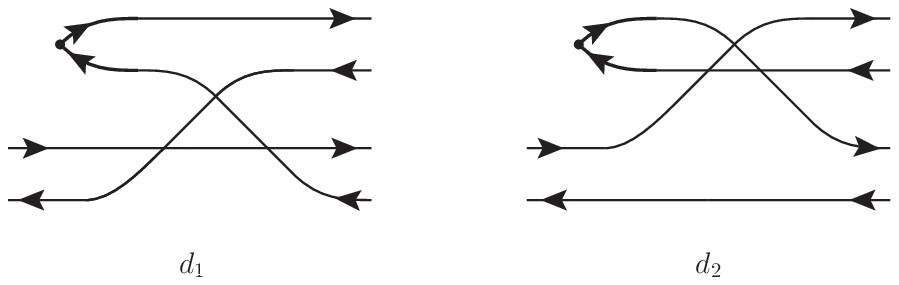}
  \caption{Decay diagrams in the $^3P_0$ model.}
  \label{decaypics}
\end{figure}

We calculate strong decay widths for the states above $B\overline{B}$ threshold using $^3P_0$ model. In the $^3P_0$ model, the open-flavor strong decay of a meson $(A\rightarrow B+C)$ take place through the production of quark anti-quark pair with vacuum quantum numbers ($J^{PC}=0^{++}$)~\cite{micu-1969}. The produced quark anti-quark pair combines with the quark anti-quark of initial meson $A$ to gives the final mesons $B$ and $C$. The interaction Hamiltonian for the $^3P_0$ model in nonrelativistic limit is

\begin{equation}\label{hamiltonian}
H_I=2 m_q \gamma\int d^3\textbf{x}\;\overline{\psi}_q(\textbf{x}) \psi_q(\textbf{x}),
\end{equation}
where
$\psi$ is the Dirac quark field and $\gamma$ is the pair-production strength parameter. We use $\gamma = 0.33$ that obtained from a fit of experimentally known strong decay widths of bottomonium states. The quark anti-quark pair production takes place through $b^{\dag}d^{\dag}$ term in the Hamiltonian
\begin{equation}\label{interaction-hamiltonian}
  H_I=2m_q\gamma \int d^3k[\overline{u}(\mathbf{k},s)v(\mathbf{-k},\overline{s})]b^{\dag}(\textbf{k},s)d^{\dag}(-\textbf{k},\overline{s}),
\end{equation}
where $b^{\dag}$ and $d^{\dag}$ are the creation operators for quark and antiquark respectively. This interaction Hamiltonian is used to calculate the matrix element $\langle BC|H_I|A\rangle$ for a process $A\rightarrow B+C$. There are two diagrams contribute in the matrix element, shown in Fig. (\ref{decaypics}).
The flavor factors for each diagram along with multiplicity factor $\mathcal{F}$ for all the processes discussed in this work are reported in Table\ref{flavor factor}.
The combined matrix element of both diagrams gives the decay amplitude
\begin{equation}\label{decay-amplitude}
  \mathcal{M}_{LS}=\langle j_A,L_{BC},S_{BC}|BC\rangle \langle BC|H_I|A\rangle/\delta(\mathbf{A}-\mathbf{B}-\mathbf{C}).
\end{equation}
The decay width of the process $A \rightarrow B+C$ can be calculated by combining the decay amplitude $(\mathcal{M}_{LS})$ with a relativistic phase space as~\cite{ackleh-1996}
\begin{equation}\label{gammaTOT}
\Gamma_{A\rightarrow BC}=2\pi\frac{P E_BE_C}{M_A}\sum_{LS} |\mathcal{M}_{LS}|^2,
\end{equation}
where $P=|\textbf{B}|=|\textbf{C}|$ in the center-of-mass of the initial meson-A, $M_A$ is the mass of this initial meson, $E_B$ and $E_C$ are the energies of the final mesons $B$ and $C$ respectively. We use experimental masses of mesons if available; otherwise our theoretically calculated masses of mesons from Table\ref{bottomonium masses} are used. The masses of the final state mesons $B$ and $B_s$ are reported in Table~\ref{B and Bs}. The detailed formulism to calculate the strong decay amplitude by using the $^3P_0$ model is described in our earlier work \cite{ishrat-2018,ishrat-2019}.\\
In this work, we have computed strong decay widths of kinematically allowed open-flavor decay modes of all the bottomonium states mentioned in Table 1 using the $^3P_0$ model.
We use simple harmonic oscillator (SHO) wavefunctions as wavefunctions of initial and final mesons in the momentum space calculations of matrix element $\langle BC|H_I|A\rangle$. The SHO scale $\beta$ for initial and final mesons is taken as parameter of the $^3P_0$ model. In this paper, we fit $\beta$ parameter of SHO wavefunctions to the numerical wavefunctions obtained by solving radial Schr\"{o}dinger equation. Our fitted $\beta$ values for the initial bottomonium mesons are reported in column-5 of Table \ref{bottomonium masses}. The $\beta$ values for the final $B$ and $B_s$ mesons appearing in strong decays of higher states of bottomonium mesons are mentioned in Table~\ref{B and Bs}.

\begin{table}[h!]
\centering
\begin{tabular}{c c c c c }
\hline
\hline
Generic Decay&Subprocess&$I_{flavor}(d_1)$&$I_{flavor}(d_2)$&$\mathcal{F}$\\
\hline
$X_b\rightarrow B \bar{B} $        &  $X_b\rightarrow B^+ B^-$            & 0 & 1  & 2  \\
$X_b\rightarrow B^* \bar{B} $      &  $X_b\rightarrow B^{*+} B^-$         & 0 & 1  & 4  \\
$X_b\rightarrow B^* \bar{B}^* $    &  $X_b\rightarrow B^{*+} B^{*-}$      & 0 & 1  & 2  \\
$X_b\rightarrow B_s \bar{B}_s $    &  $X_b\rightarrow B_s^+ B_s^-$        & 0 & 1  & 1  \\
$X_b\rightarrow B_s^*\bar{B}_s $   &  $X_b\rightarrow B_s^{*+} B_s^-$     & 0 & 1  & 2  \\
$X_b\rightarrow B_s^*\bar{B}_s^* $ &  $X_b\rightarrow B_s^{*+} B_s^{*-}$  & 0 & 1  & 1  \\
\hline
\end{tabular}
\caption{Flavor factors for bottomonium decay, where $|X_b\rangle\equiv|b\bar{b}\rangle$.}\label{flavor factor}
\end{table}

\begin{table}[h!]
\centering
\begin{tabular}{c c c c c }
\hline
\hline
Meson           & state         &Expt. Mass(GeV) &Our calculated Mass(GeV)& $\beta$ (GeV)   \\
\hline$B$       &$1^1S_0$       &5.279                        &5.2553                      &0.450        \\
$B^*$           &$1^3S_1$       &5.325                        &5.2597                      &0.448        \\
$B_s$           &$1^1S_0$       &5.367                        &5.3892                      &0.440       \\
$B_s^*$         &$1^3S_1$       &5.415                        &5.3926                      &0.439        \\
\hline
\end{tabular}
\caption{Masses and SHO $\beta$ values of B and $B_s$ mesons used in calculations of strong decay widths.}\label{B and Bs}
\end{table}

\begin{table}[h!]
\centering
\begin{tabular}{c c c c c c}
\hline
\hline
Initial State&Final State&Predicted Width(KeV)&Predicted B.R($\%$)&Measured Width(KeV)~\cite{PDG-22}\\
\hline\hline
$\Upsilon (1 ^3S_1) $        &  $\eta_b(1 ^1S_0)\gamma$   & $3.1979 \times 10^{-6}$ & 100  & $54.02\pm1.25$  \\
\hline
$\eta_b (2 ^1S_0) $         &  $h_b(1 ^1P_1)\gamma$   &  1.5892& 99.99  &   \\
                            &  $\Upsilon(1 ^3S_1)\gamma$   & 0.000075999 & 0.0048 &   \\
                            &Total& 1.58928 & 100 & \\
\hline
$\Upsilon (2 ^3S_1) $       &  $\chi_{b2}(1 ^3P_2)\gamma$   & 0.7283  & 44.97  &   \\
                            &  $\chi_{b1}(1 ^3P_1)\gamma$   & 0.5854 & 36.15  &   \\
                            &  $\chi_{b0}(1 ^3P_0)\gamma$   & 0.3055 &  18.86 &   \\
                            &  $\eta_{b}(2 ^1S_0)\gamma$   & $8.0239 \times 10^{-8}$ &  $4.95 \times 10^{-6}$ &   \\
                            &  $\eta_{b}(1 ^1S_0)\gamma$   & 0.00002618 & 0.0016  &   \\
                            &Total& 1.619 &100& $31.98\pm2.63$\\
\hline
$\eta_b (3 ^1S_0) $         &  $h_b(2 ^1P_1)\gamma$   & 2.90075 & 88.45  &   \\
                            &  $h_b(1 ^1P_1)\gamma$   &  0.3785& 11.54  &   \\
                            &  $\Upsilon(2 ^3S_1)\gamma$   & 0.00002351 & 0.0007  &   \\
                            &  $\Upsilon(1 ^3S_1)\gamma$   & 0.00003586 & 0.0011  &   \\
                            &Total& 3.279 & 100\\
\hline
$\Upsilon (3 ^3S_1) $       &  $\chi_{b2}(2 ^3P_2)\gamma$   & 1.3933 & 41.5   &   \\
                            &  $\chi_{b1}(2 ^3P_1)\gamma$   & 1.0615 & 31.62  &   \\
                            &  $\chi_{b0}(2 ^3P_0)\gamma$   & 0.5111 & 15.22  &   \\
                            &  $\chi_{b2}(1 ^3P_2)\gamma$   & 0.2087 & 6.22  &   \\
                            &  $\chi_{b1}(1 ^3P_1)\gamma$   & 0.1332 & 3.97  &   \\
                            &  $\chi_{b0}(1 ^3P_0)\gamma$   & 0.04931 & 1.47  &   \\
                            &  $\eta_{b}(3 ^1S_0)\gamma$   & $2.3777 \times 10^{-8}$ &$7.08 \times 10^{-7}$ &   \\
                            &  $\eta_{b}(2 ^1S_0)\gamma$   & $7.9787 \times 10^{-6}$ & 0.0002  &   \\
                            &  $\eta_{b}(1 ^1S_0)\gamma$   & $0.1238 \times 10^{-4}$ &  0.0003 &   \\
                            &Total& 3.357 & 100& $20.32\pm1.85$\\
\hline
\end{tabular}
\caption{Partial widths of radiative transitions and strong decays for 1S, 2S and 3S bottomonium mesons.}\label{results-1}
\end{table}

\begin{table}[h!]
\centering
\begin{tabular}{c c c c c c}
\hline
\hline
Initial State&Final State&Predicted Width(KeV)&Predicted B.R($\%$)&Measured Width(KeV)~\cite{PDG-22}\\
\hline\hline
$\eta_b (4 ^1S_0) $         &  $h_b(3 ^1P_1)\gamma$   & 0.39029 & 39.51  &   \\
                            &  $h_b(2 ^1P_1)\gamma$   & 0.3761 & 38.08  &   \\
                            &  $h_b(1 ^1P_1)\gamma$   & 0.2213 & 22.4  &   \\
                            &  $\Upsilon(3 ^3S_1)\gamma$   & 0.000010108 & 0.001  &   \\
                            &  $\Upsilon(2 ^3S_1)\gamma$   & 0.000022665 & 0.002  &   \\
                            &  $\Upsilon(1 ^3S_1)\gamma$   & 0.000019219 & 0.002  &   \\
                            &Total& 0.9877 & 100\\
\hline
$\Upsilon (4 ^3S_1) $       &  $\chi_{b2}(3 ^3P_2)\gamma$   & 1.8892 & 0.009  &   \\
                            &  $\chi_{b1}(3 ^3P_1)\gamma$   & 1.4169 & 0.007  &   \\
                            &  $\chi_{b0}(3 ^3P_0)\gamma$   & 0.6644 & 0.003  &   \\
                            &  $\chi_{b2}(2 ^3P_2)\gamma$   & 0.2055 & 0.001  &   \\
                            &  $\chi_{b1}(2 ^3P_1)\gamma$   & 0.1304 & 0.0006  &   \\
                            &  $\chi_{b0}(2 ^3P_0)\gamma$   & 0.04769 & 0.0002  &   \\
                            &  $\chi_{b2}(1 ^3P_2)\gamma$   & 0.1229 & 0.0006  &   \\
                            &  $\chi_{b1}(1 ^3P_1)\gamma$   & 0.07654 & 0.0004  &   \\
                            &  $\chi_{b0}(1 ^3P_0)\gamma$   & 0.02720 & 0.0001  &   \\
                            &  $\eta_{b}(1 ^1S_0)\gamma$   & $6.6468  \times 10^{-6}$ & $3.22  \times 10^{-8}$  &   \\
                            &  $BB$   & 20.64 MeV &  99.98 & \\
                            &Total& 20.645 MeV & 100 & $20.5\pm2.5$ MeV \\
\hline
\end{tabular}
\caption{Partial widths of radiative transitions and strong decays for 4S bottomonium mesons.}\label{results-2}
\end{table}

\begin{table}[h!]
\centering
\begin{tabular}{c c c c c c}
\hline
\hline
Initial State&Final State&Predicted Width(KeV)&Predicted B.R($\%$)&Measured Width(KeV)~\cite{PDG-22}\\
\hline\hline
$\eta_b (5 ^1S_0) $        &  $h_b(4 ^1P_1)\gamma$   & 3.5609 &  0.007 &   \\
                             &  $h_b(3 ^1P_1)\gamma$   & 0.305445 & 0.0006  &   \\
                             &  $h_b(2 ^1P_1)\gamma$   & 0.2418 & 0.0005  &   \\
                             &  $h_b(1 ^1P_1)\gamma$   & 0.1339 & 0.0003  &   \\
                             &  $\Upsilon(4 ^3S_1)\gamma$   & $5.3135 \times 10^{-6}$ & $1.005 \times 10^{-8}$  &   \\
                             &  $\Upsilon(3 ^3S_1)\gamma$   & $1.182 \times 10^{-5}$  & $2.23 \times 10^{-8}$   &   \\
                             &  $\Upsilon(2 ^3S_1)\gamma$   & $1.708 \times 10^{-5}$   &  $3.23 \times 10^{-8}$  &   \\
                             &  $\Upsilon(1 ^3S_1)\gamma$   & $1.0928 \times 10^{-5}$  &  $2.07 \times 10^{-8}$ &   \\
                             &  $BB^*$   & 28.74 MeV & 54.33  &  &   \\
                             &  $B^*B^*$   & 13.45 MeV &  25.43 &  &   \\
                             &  $B_sB_s^*$   & 8.13 MeV & 15.37  &  &   \\
                             &  $B_s^*B_s^*$   & 2.57 MeV & 4.86  &  &   \\
                             & Total & 52.894 MeV & 100 & \\
\hline
$\Upsilon (5 ^3S_1) $       &  $\chi_{b2}(4 ^3P_2)\gamma$   &1.7238  & 0.003  &   \\
                            &  $\chi_{b1}(4 ^3P_1)\gamma$   &1.2884  & 0.0026  &   \\
                            &  $\chi_{b0}(4 ^3P_0)\gamma$   & 0.6005 & 0.001  &   \\
                            &  $\chi_{b2}(3 ^3P_2)\gamma$   & 0.1661 & 0.0003  &   \\
                            &  $\chi_{b1}(3 ^3P_1)\gamma$   &0.1055  & 0.0002  &   \\
                            &  $\chi_{b0}(3 ^3P_0)\gamma$   &0.0386  &  0.0001 &   \\
                            &  $\chi_{b2}(2 ^3P_2)\gamma$   & 0.1339 & 0.0003  &   \\
                            &  $\chi_{b1}(2 ^3P_1)\gamma$   & 0.0831 & 0.0002  &   \\
                            &  $\chi_{b0}(2 ^3P_0)\gamma$   & 0.0293 & 0.0001  &   \\
                            &  $\chi_{b2}(1 ^3P_2)\gamma$   & 0.0744 & 0.0001  &   \\
                            &  $\chi_{b1}(1 ^3P_1)\gamma$   & 0.0459 & 0.0001  &   \\
                            &  $\chi_{b0}(1 ^3P_0)\gamma$   & 0.016 & 0.00003  &   \\
                            &  $\eta_{b}(5 ^1S_0)\gamma$   & $2.9724 \times 10^{-9}$  & $5.89 \times 10^{-12}$  &   \\
                            &  $\eta_{b}(4 ^1S_0)\gamma$   & $1.6909 \times 10^{-6}$  & $3.35 \times 10^{-9}$  &   \\
                            &  $\eta_{b}(3 ^1S_0)\gamma$   & $3.9915 \times 10^{-6}$   & $7.91 \times 10^{-9}$   &   \\
                            &  $\eta_{b}(2 ^1S_0)\gamma$   & $5.8163 \times 10^{-6}$  & $1.15 \times 10^{-8}$  &   \\
                            &  $\eta_{b}(1 ^1S_0)\gamma$   & $3.784 \times 10^{-6}$   &  $7.5 \times 10^{-9}$  &   \\
                            &  $BB$   & 3.11 MeV &  6.16 &  &   \\
                             &  $BB^*$   & 19.57 MeV& 38.77  &  &   \\
                             &  $B^*B^*$   & 20.5 MeV & 40.61  &  &   \\
                             &  $B_sB_s$   & 0.57 MeV &  1.13 &  &   \\
                             &  $B_sB_s^*$   & 6.42 MeV&  12.72 &  &   \\
                             &  $B_s^*B_s^*$   & 0.3 MeV & 0.6  &  &   \\
                             & Total & 50.47 MeV & 100 & $37\pm4$ MeV&\\
\hline
\end{tabular}
\caption{Partial widths of radiative transitions and strong decays for 5S bottomonium mesons.}\label{results-3}
\end{table}

\begin{table}[h!]
\centering
\begin{tabular}{c c c c c c}
\hline
\hline
Initial State&Final State&Predicted Width(KeV)&Predicted B.R($\%$)\\
\hline\hline
$h_b (1 ^1P_1) $            &  $\chi_{b0}(1 ^3P_0)\gamma$   & $ 0.3469 \times 10^{-4}$ & 0.0002   \\
                            &  $\chi_{b1}(1 ^3P_1)\gamma$   & $6.1806 \times 10^{-8}$ & $2.9 \times 10^{-7}$    \\
                            &  $\eta_b(1 ^1S_0)\gamma$   & 21.3058 & 99.99    \\
                            &Total& 21.3058 & 100\\
\hline
$\chi_{b0} (1 ^3P_0) $      &  $\Upsilon(1 ^3S_1)\gamma$   & 18.5847 & 100     \\
\hline
$\chi_{b1} (1 ^3P_1) $      &  $\Upsilon(1 ^3S_1)\gamma$   & 20.5551 & 100     \\
\hline
$\chi_{b2} (1 ^3P_2) $      &  $h_b(1 ^1P_1)\gamma$   &$9.0494 \times 10^{-6}$  & 0.00004    \\
                            &  $\Upsilon(1 ^3S_1)\gamma$   & 21.7469 & $\sim 100$   \\
                            &Total& 21.7469\\
\hline
\end{tabular}
\caption{Partial widths of radiative transitions and strong decays for 1P bottomonium mesons.}\label{results-4}
\end{table}

\begin{table}[h!]
\centering
\begin{tabular}{c c c c c c}
\hline
\hline
Initial State&Final State&Predicted Width(KeV)&Predicted B.R($\%$)\\
\hline\hline
$h_b (2 ^1P_1) $            &  $\eta_{b2}(1 ^1D_2)\gamma$   & 1.6063 & 8.32    \\
                            &  $\chi_{b0}(2 ^3P_0)\gamma$   & $0.1499 \times 10^{-4}$ & 0.0001     \\
                            &  $\chi_{b1}(2 ^3P_1)\gamma$   & $ 8.0239 \times 10^{-8}$ & $ 4.15 \times 10^{-7}$    \\
                            &  $\chi_{b0}(1 ^3P_0)\gamma$   & $4.3222 \times 10^{-6}$ &  0.00002    \\
                            &  $\chi_{b1}(1 ^3P_1)\gamma$   & $ 0.1138 \times 10^{-4}$ & 0.0001      \\
                            &  $\chi_{b2}(1 ^3P_2)\gamma$   & $0.1754 \times 10^{-4}$  & 0.0001     \\
                            &  $\eta_b(2 ^1S_0)\gamma$   & 10.1237  &  52.46   \\
                            &  $\eta_b(1 ^1S_0)\gamma$   & 7.5683 &  39.22   \\
                            &Total&19.298 & 100 \\
\hline
$\chi_{b0} (2 ^3P_0) $      &  $\Upsilon_1(1 ^3D_1)\gamma$   & 7.1356 & 42.38     \\
                            &  $h_b(1 ^1P_1)\gamma$   & $0.1 \times 10^{-4}$ &0.0001     \\
                            &  $\Upsilon(2 ^3S_1)\gamma$   & 8.4946 & 50.45     \\
                            &  $\Upsilon(1 ^3S_1)\gamma$   & 1.2087 &  7.178   \\
                            &Total& 16.8389 & 100\\
\hline
$\chi_{b1} (2 ^3P_1) $      &  $\Upsilon_2(1 ^3D_2)\gamma$   & 1.1361 & 6.05     \\
                            &  $\Upsilon_1(1 ^3D_1)\gamma$   &0.4261  & 2.27     \\
                            &  $h_b(1 ^1P_1)\gamma$   & $0.1105 \times 10^{-4}$  & 0.0001     \\
                            &  $\Upsilon(2 ^3S_1)\gamma$   & 9.7699 & 52.06     \\
                            &  $\Upsilon(1 ^3S_1)\gamma$   & 7.4319 & 39.61     \\
                            &Total& 18.764 & 100\\
\hline
$\chi_{b2} (2 ^3P_2) $      &  $\Upsilon_3(1 ^3D_3)\gamma$   & 1.4371  & 7.22     \\
                            &  $\Upsilon_2(1 ^3D_2)\gamma$   & 0.2752 & 1.38     \\
                            &  $\Upsilon_1(1 ^3D_1)\gamma$   & 0.0205 & 0.1     \\
                            &  $h_b(2 ^1P_1)\gamma$   & $ 2.7562 \times 10^{-6}$ &0.00001    \\
                            &  $h_b(1 ^1P_1)\gamma$   & $ 0.117 \times 10^{-4}$  & 0.0001    \\
                            &  $\Upsilon(2 ^3S_1)\gamma$   & 10.5728 & 53.09     \\
                            &  $\Upsilon(1 ^3S_1)\gamma$   & 7.6091 & 38.21     \\
                            &Total& 7.2163 & 100\\
\hline
\end{tabular}
\caption{Partial widths of radiative transitions and strong decays for 2P bottomonium mesons.}\label{results-5}
\end{table}

\begin{table}[h!]
\centering
\begin{tabular}{c c c c c c}
\hline
\hline
Initial State&Final State&Predicted Width(KeV)&Predicted B.R($\%$)\\
\hline\hline
$h_b (3 ^1P_1) $            &  $\eta_{b2}(2 ^1D_2)\gamma$   & 2.7351  & 14.34   \\
                            &  $\eta_{b2}(1 ^1D_2)\gamma$   & 0.004368 & 0.023     \\
                            &  $\chi_{b0}(3 ^3P_0)\gamma$   & $9.8275 \times 10^{-6}$ & 0.0001     \\
                            &  $\chi_{b1}(3 ^3P_1)\gamma$   & $8.024 \times 10^{-8}$ & $4.2 \times 10^{-7}$    \\
                            &  $\chi_{b0}(2 ^3P_0)\gamma$   & $1.6508 \times 10^{-6}$ & $8.66 \times 10^{-6}$      \\
                            &  $\chi_{b1}(2 ^3P_1)\gamma$   & $4.3966 \times 10^{-6}$ & 0.00002    \\
                            &  $\chi_{b2}(2 ^3P_2)\gamma$   & $6.8177 \times 10^{-6}$ & 0.00004     \\
                            &  $\chi_{b0}(1 ^3P_0)\gamma$   & $3.0805 \times 10^{-6}$ & 0.00002     \\
                            &  $\chi_{b1}(1 ^3P_1)\gamma$   & $8.5979 \times 10^{-4}$  & 0.005     \\
                            &  $\chi_{b2}(1 ^3P_2)\gamma$   & $0.1374 \times 10^{-4}$ & 0.0001     \\
                            &  $\eta_b(3 ^1S_0)\gamma$   & 8.6711 & 45.47     \\
                            &  $\eta_b(2 ^1S_0)\gamma$   & 3.329 &  17.46    \\
                            &  $\eta_b(1 ^1S_0)\gamma$   & 4.3278 & 22.7     \\
                            &Total& 19.0683 &  100\\
\hline
$\chi_{b0} (3 ^3P_0) $      &  $\Upsilon_1(2 ^3D_1)\gamma$   & 2.1021 &12.75     \\
                            &  $\Upsilon_1(1 ^3D_1)\gamma$   & $0.3958 \times 10^{-2}$ & 0.024    \\
                            &  $h_b(2 ^1P_1)\gamma$   & $3.7949 \times 10^{-6}$  & 0.00002  \\
                            &  $h_b(1 ^1P_1)\gamma$   & $8.2054 \times 10^{-6}$ &  0.00005    \\
                            &  $\Upsilon(3 ^3S_1)\gamma$   & 7.1219 & 43.21     \\
                            &  $\Upsilon(2 ^3S_1)\gamma$   & 3.0933 & 18.77     \\
                            &  $\Upsilon(1 ^3S_1)\gamma$   & 4.1616 & 25.25     \\
                            &Total& 16.4829 &100\\
\hline
$\chi_{b1} (3 ^3P_1) $      &  $\Upsilon_2(2 ^3D_2)\gamma$   & 1.9323 & 10.43     \\
                            &  $\Upsilon_1(2 ^3D_1)\gamma$   & 0.7243 & 3.91     \\
                            &  $\Upsilon_2(1 ^3D_2)\gamma$   & $0.3103 \times 10^{-2}$ & 0.017     \\
                            &  $\Upsilon_1(1 ^3D_1)\gamma$   & $0.1068 \times 10^{-2}$ & 0.006     \\
                            &  $h_b(2 ^1P_1)\gamma$   & $4.2216 \times 10^{-6}$ &  0.00002    \\
                            &  $h_b(1 ^1P_1)\gamma$ & $8.5899 \times 10^{-6}$ &  0.00005   \\
                            &  $\Upsilon(3 ^3S_1)\gamma$   & 8.3322 & 44.97     \\
                            &  $\Upsilon(2 ^3S_1)\gamma$   & 3.2667 & 17.63     \\
                            &  $\Upsilon(1 ^3S_1)\gamma$   & 4.2684 & 23.04     \\
                            &Total& 18.5281 &100 \\
\hline
$\chi_{b2} (3 ^3P_2) $      &  $\Upsilon_3(2 ^3D_3)\gamma$   & 2.4236 &  12.27    \\
                            &  $\Upsilon_2(2 ^3D_2)\gamma$   & 0.4638 & 2.35     \\
                            &  $\Upsilon_1(2 ^3D_1)\gamma$   & 0.0345 & 0.175    \\
                            &  $\Upsilon_3(1 ^3D_3)\gamma$   & $0.3565 \times 10^{-2}$ & 0.018    \\
                            &  $\Upsilon_2(1 ^3D_2)\gamma$   & $0.649 \times 10^{-3}$ & 0.003     \\
                            &  $\Upsilon_1(1 ^3D_1)\gamma$   & $0.4463 \times 10^{-4}$ & 0.0002     \\
                            &  $h_b(3 ^1P_1)\gamma$   & $1.5208 \times 10^{-6}$ &  $7.7 \times 10^{-6}$    \\
                            &  $h_b(2 ^1P_1)\gamma$   & $4.4896 \times 10^{-6}$ &  0.00002    \\
                            &  $h_b(1 ^1P_1)\gamma$   & $0.8239 \times 10^{-6}$ & $4.12 \times 10^{-6}$     \\
                            &  $\Upsilon(3 ^3S_1)\gamma$   & 9.1123 &  46.15    \\
                            &  $\Upsilon(2 ^3S_1)\gamma$   & 3.3727 & 17.08     \\
                            &  $\Upsilon(1 ^3S_1)\gamma$   & 4.3327 &  21.94    \\
                            &Total& 19.7439 & 100\\
\hline
\end{tabular}
\caption{Partial widths of radiative transitions and strong decays for 3P bottomonium mesons.}\label{results-6}
\end{table}

\begin{table}[h!]
\centering
\begin{tabular}{c c c c c c}
\hline
\hline
Initial State&Final State&Predicted Width(KeV)&Predicted B.R($\%$)\\
\hline\hline
$\eta_{b2} (1 ^1D_2) $      &  $h_b(1 ^1P_1)\gamma$   & 13.8754 &  $\sim 100$ &   \\
                            &  $\Upsilon_1 (1 ^3D_1)\gamma$   & $6.1137 \times 10^{-7}$ &$4.4 \times 10^{-6}$  \\
                            &  $\Upsilon_2 (1 ^3D_2)\gamma$   & $1.2542 \times 10^{-9}$ &  $8.9 \times 10^{-9}$    \\
                            &Total& 13.8754 & 100\\
\hline
$\Upsilon_1 (1 ^3D_1) $     &  $\chi_{b0}(1 ^3P_0)\gamma$   & 9.1119 & 60.31     \\
                            &  $\chi_{b1}(1 ^3P_1)\gamma$   & 5.6607 & 37.47     \\
                            &  $\chi_{b2}(1 ^3P_2)\gamma$   & 0.3367 &  2.23    \\
                            &Total&15.1093 & 100 \\
\hline
$\Upsilon_2 (1 ^3D_2) $     &  $\chi_{b1}(1 ^3P_1)\gamma$   & 10.7101 & 77.04     \\
                            &  $\chi_{b2}(1 ^3P_2)\gamma$   & 3.1914 & 22.96     \\
                            &Total& 13.9015 & 100\\
\hline
$\Upsilon_3 (1 ^3D_3) $     &  $\chi_{b2}(1 ^3P_2)\gamma$   & 13.1715 & $\sim 100$     \\
                            &  $\eta_{b2}(1 ^1D_2)\gamma$   & $4.9432 \times 10^{-7}$ & $3.7\times 10^{-6}$     \\
                            &Total& 13.1715 & 100\\
\hline
\end{tabular}
\caption{Partial widths of radiative transitions and strong decays for 1D bottomonium mesons.}\label{results-7}
\end{table}

\begin{table}[h!]
\centering
\begin{tabular}{c c c c c c}
\hline
\hline
Initial State&Final State&Predicted Width(KeV)&Predicted B.R($\%$)\\
\hline\hline
$\eta_{b2} (2 ^1D_2) $      &  $h_b(2 ^1P_1)\gamma$   & 10.8722 & 75.61     \\
                            &  $h_b(1 ^1P_1)\gamma$   & 2.0873 & 14.53    \\
                            &  $\Upsilon_1 (2 ^3D_1)\gamma$   &$4.8952 \times 10^{-7}$  & $3.4 \times 10^{-6}$    \\
                            &  $\Upsilon_1 (1 ^3D_1)\gamma$   & $1.5098 \times 10^{-6}$ & 0.00001     \\
                            &  $\Upsilon_2 (1 ^3D_2)\gamma$   & $2.4132 \times 10^{-6}$ &  0.00002    \\
                            &  $\Upsilon_3 (1 ^3D_3)\gamma$   & $3.2919 \times 10^{-6}$ & 0.00002     \\
                            &  $h_{b3} (1 ^1F_3)\gamma$   & 1.4080 & 9.8    \\
                            &Total& 14.3675 & 100\\
\hline
$\Upsilon_1 (2 ^3D_1) $     &  $\chi_{b0}(2 ^3P_0)\gamma$   & 7.0574 &  46.55   \\
                            &  $\chi_{b1}(2 ^3P_1)\gamma$   & 4.4115 &  29.1    \\
                            &  $\chi_{b2}(2 ^3P_2)\gamma$   & 0.2628 &  1.73    \\
                            &  $\chi_{b0}(1 ^3P_0)\gamma$   &1.2310  &  8.12  \\
                            &  $\chi_{b1}(1 ^3P_1)\gamma$   & 0.8473 &  5.59    \\
                            &  $\chi_{b2}(1 ^3P_2)\gamma$   & 0.0537 & 0.35     \\
                            &  $\eta_{b2} (1 ^1D_2)\gamma$   & $2.3319 \times 10^{-6}$ & 0.00002    \\
                            &  $\chi_{b2} (1 ^3F_2)\gamma$   & 1.2964 & 8.55     \\
                            &Total& 15.1601 &100 \\
\hline
$\Upsilon_2 (2 ^3D_2) $     &  $\chi_{b1}(2 ^3P_1)\gamma$   & 8.4307 &  58.55   \\
                            &  $\chi_{b2}(2 ^3P_2)\gamma$   & 2.5175 &  17.49    \\
                            &  $\chi_{b1}(1 ^3P_1)\gamma$   & 1.5564 &  10.81    \\
                            &  $\chi_{b2}(1 ^3P_2)\gamma$   & 0.4934 &  3.43    \\
                            &  $\eta_{b2}(1 ^1D_2)\gamma$   & $2.4228 \times 10^{-6}$ & 0.00002    \\
                            &  $\chi_{b2} (1 ^3F_2)\gamma$   & 0.1610 & 1.12     \\
                            &  $\chi_{b3} (1 ^3F_3)\gamma$   & 1.2387 &  8.6   \\
                            &Total& 14.3977 & 100\\
\hline
$\Upsilon_3 (2 ^3D_3) $     &  $\chi_{b2}(2 ^3P_2)\gamma$   & 10.451 &  75.01 &   \\
                            &  $\chi_{b2}(1 ^3P_2)\gamma$   & 1.9983 &  14.34 &   \\
                            &  $\eta_{b2}(2 ^1D_2)\gamma$   & $3.1845 \times 10^{-7}$ &$2.28 \times 10^{-6}$   &   \\
                            &  $\eta_{b2}(1 ^1D_2)\gamma$   &$2.4792 \times 10^{-6}$  & 0.00002  &   \\
                            &  $\chi_{b2} (1 ^3F_2)\gamma$   & $0.3510 \times 10^{-2}$ & 0.025  &   \\
                            &  $\chi_{b3} (1 ^3F_3)\gamma$   & 0.1182 & 0.85  &   \\
                            &  $\chi_{b4} (1 ^3F_4)\gamma$   & 1.3627 & 9.78  &   \\
                            &Total& 13.9337 & 100\\
\hline
\end{tabular}
\caption{Partial widths of radiative transitions and strong decays for 2D bottomonium mesons.}\label{results-8}
\end{table}

\begin{table}[h!]
\centering
\begin{tabular}{c c c c c c}
\hline
\hline
Initial State&Final State&Predicted Width(KeV)&Predicted B.R($\%$)\\
\hline\hline
$\eta_{b2} (3 ^1D_2) $      &  $h_b(3 ^1P_1)\gamma$   & 9.6861 &  66.21 &   \\
                            &  $h_b(2 ^1P_1)\gamma$   & 1.7884 & 12.22  &   \\
                            &  $h_b(1 ^1P_1)\gamma$   & 0.8040 &  5.5 &   \\
                            &  $h_{b3} (2 ^1F_3)\gamma$   & 2.3502 & 16.06  &   \\
                            &  $h_{b3} (1 ^1F_3)\gamma$   & $0.4152 \times 10^{-3}$ & 0.003  &   \\
                            &  $\Upsilon_1(1 ^3 D_1)\gamma$   & $1.617 \times 10^{-6}$ & 0.00001  &   \\
                            &  $\Upsilon_2(1 ^3 D_2)\gamma$   & $2.6358 \times 10^{-6}$ & 0.00002  &   \\
                            &  $\Upsilon_3(1 ^3 D_3)\gamma$   & $3.6399 \times 10^{-6}$ & 0.00002  &   \\
                            &Total& 14.6291& 100 & \\
\hline
$\Upsilon_1 (3 ^3D_1) $     &  $\chi_{b0}(3 ^3P_0)\gamma$   & 6.2996 & 41.24  &   \\
                            &  $\chi_{b1}(3 ^3P_1)\gamma$   & 3.9224 & 25.68  &   \\
                            &  $\chi_{b2}(3 ^3P_2)\gamma$   & 0.2327 &  1.52 &   \\
                            &  $\chi_{b0}(2 ^3P_0)\gamma$   &1.0527  & 6.89  &   \\
                            &  $\chi_{b1}(2 ^3P_1)\gamma$   & 0.7308 & 4.78  &   \\
                            &  $\chi_{b2}(2 ^3P_2)\gamma$   & 0.04651 & 0.304  &   \\
                            &  $\chi_{b0}(1 ^3P_0)\gamma$   & 0.4607 & 3.02  &   \\
                            &  $\chi_{b1}(1 ^3P_1)\gamma$   & 0.3261 &  2.13 &   \\
                            &  $\chi_{b2}(1 ^3P_2)\gamma$   & 0.02102 &  0.137 &   \\
                            &  $\eta_{b2} (1 ^1D_2)\gamma$   &$2.6154  \times 10^{-6}$ & 0.00002  &   \\
                            &  $\chi_{b2} (2 ^3F_2)\gamma$   & 2.1778 & 14.26  &   \\
                            &  $\chi_{b2} (1 ^3F_2)\gamma$   & $5.5787 \times 10^{-3}$& 0.037  &   \\
                            &Total&15.2759 & 100 & \\
\hline
$\Upsilon_2 (3 ^3D_2) $     &  $\chi_{b1}(3 ^3P_1)\gamma$   & 7.5476 & 51.33  &   \\
                            &  $\chi_{b2}(3 ^3P_2)\gamma$   & 2.2451 & 15.27  &   \\
                            &  $\chi_{b1}(2 ^3P_1)\gamma$   & 1.3464 &  9.16 &   \\
                            &  $\chi_{b2}(2 ^3P_2)\gamma$   & 0.4287 & 2.92  &   \\
                            &  $\chi_{b1}(1 ^3P_1)\gamma$   & 0.5945 &  4.04 &   \\
                            &  $\chi_{b2}(1 ^3P_2)\gamma$   & 0.1916 & 1.3  &   \\
                            &  $\eta_{b2}(2 ^1D_2)\gamma$   & $1.5963 \times 10^{-6}$ & 0.00001  &   \\
                            &  $\eta_{b2}(1 ^1D_2)\gamma$   &  $2.6649 \times 10^{-6}$&  0.00002 &   \\
                            &  $\chi_{b2} (2 ^3F_2)\gamma$   & 0.2710 & 1.84  &   \\
                            &  $\chi_{b3} (2 ^3F_3)\gamma$   & 2.0766 & 14.12  &   \\
                            &Total&14.7015 & 100 & \\
\hline
$\Upsilon_3 (3 ^3D_3) $     &  $\chi_{b2}(3 ^3P_2)\gamma$   & 9.3650 &  65.26 &   \\
                            &  $\chi_{b2}(2 ^3P_2)\gamma$   & 1.7397 &  12.12 &   \\
                            &  $\chi_{b2}(1 ^3P_2)\gamma$   & 0.7725 &  5.38 &   \\
                            &  $\eta_{b2}(2 ^1D_2)\gamma$   & $1.6370 \times 10^{-6}$ & 0.00001  &   \\
                            &  $\eta_{b2}(1 ^1D_2)\gamma$   & $2.6957 \times 10^{-6}$ &  0.00002 &   \\
                            &  $\chi_{b2} (2 ^3F_2)\gamma$   & $0.5917 \times 10^{-2}$ & 0.04  &   \\
                            &  $\chi_{b3} (2 ^3F_3)\gamma$   & 0.1986 &  1.38 &   \\
                            &  $\chi_{b4} (2 ^3F_4)\gamma$   & 2.2687 &  15.81 &   \\
                            &Total& 14.3504 & 100 & \\
\hline
\end{tabular}
\caption{Partial widths of radiative transitions and strong decays for 3D bottomonium mesons.}\label{results-9}
\end{table}

\begin{table}[h!]
\centering
\begin{tabular}{c c c c c c}
\hline
\hline
Initial State&Final State&Predicted Width(KeV)&Predicted B.R($\%$)\\
\hline\hline
$\eta_{b2} (4 ^1D_2) $      &  $h_b(4 ^1P_1)\gamma$   & 7.1699 & 0.15  &   \\
                            &  $h_b(3 ^1P_1)\gamma$   & 1.3936 & 0.03  &   \\
                            &  $h_b(2 ^1P_1)\gamma$   & 0.6984 & 0.014  &   \\
                            &  $h_b(1 ^1P_1)\gamma$   & 0.4044 & 0.008  &   \\
                            &  $h_{b3} (2 ^1F_3)\gamma$ & 0.001835 & 0.00004  &   \\
                            &  $h_{b3} (1 ^1F_3)\gamma$ & $2.4849 \times 10^{-4} $& $5.83 \times 10^{-6} $  &   \\
                            &  $BB^*$   & 4.83 MeV & 99.8  &   \\
                            &Total& 4.839 MeV & 100& \\
\hline
$\Upsilon_1 (4 ^3D_1) $     &  $\chi_{b0}(4 ^3P_0)\gamma$   & 4.6729 & 0.14  &   \\
                            &  $\chi_{b1}(4 ^3P_1)\gamma$   & 2.8896 & 0.09  &   \\
                            &  $\chi_{b2}(4 ^3P_2)\gamma$   & 0.17 &  0.005 &   \\
                            &  $\chi_{b0}(3 ^3P_0)\gamma$   & 0.8242 & 0.026  &   \\
                            &  $\chi_{b1}(3 ^3P_1)\gamma$   & 0.5713 & 0.018  &   \\
                            &  $\chi_{b2}(3 ^3P_2)\gamma$   & 0.0363 & 0.001  &   \\
                            &  $\chi_{b0}(2 ^3P_0)\gamma$   & 0.4010 & 0.013  &   \\
                            &  $\chi_{b1}(2 ^3P_1)\gamma$   & 0.2856 & 0.0089  &   \\
                            &  $\chi_{b2}(2 ^3P_2)\gamma$   & 0.0185 & 0.0006  &   \\
                            &  $\chi_{b0}(1 ^3P_0)\gamma$   & 0.2291 & 0.007  &   \\
                            &  $\chi_{b1}(1 ^3P_1)\gamma$   & 0.1642 & 0.005  &   \\
                            &  $\chi_{b2}(1 ^3P_2)\gamma$   & 0.01066 & 0.0003  &   \\
                            &  $BB$                         & 2.87 MeV & 89.68& \\
                            & $BB^*$                        & 0.32 MeV & 9.99 & \\
                            &Total& 3.2 MeV & 100 & \\
\hline
$\Upsilon_2 (4 ^3D_2) $     &  $\chi_{b1}(4 ^3P_1)\gamma$   & 5.5606 & 0.16  &   \\
                            &  $\chi_{b2}(4 ^3P_2)\gamma$   & 1.6461 &  0.048 &   \\
                            &  $\chi_{b1}(3 ^3P_1)\gamma$   & 1.0518 & 0.031  &   \\
                            &  $\chi_{b2}(3 ^3P_2)\gamma$   & 0.3343 & 0.01  &   \\
                            &  $\chi_{b1}(2 ^3P_1)\gamma$   & 0.5207 & 0.015  &   \\
                            &  $\chi_{b2}(2 ^3P_2)\gamma$   & 0.1683 & 0.005  &   \\
                            &  $\chi_{b1}(1 ^3P_1)\gamma$   & 0.2980 &  0.0087 &   \\
                            &  $\chi_{b2}(1 ^3P_2)\gamma$   & 0.09676 & 0.0028  &   \\
                            &  $BB^* $                 & 3.4 MeV & 99.7 & \\
                             &Total&3.41 MeV & 100 & \\
\hline
$\Upsilon_3 (4 ^3D_3) $     &  $\chi_{b2}(4 ^3P_2)\gamma$   & 6.8685 & 0.11  &   \\
                            &  $\chi_{b2}(3 ^3P_2)\gamma$   & 1.3563 &  0.022 &   \\
                            &  $\chi_{b2}(2 ^3P_2)\gamma$   & 0.6786 & 0.011  &   \\
                            &  $\chi_{b2}(1 ^3P_2)\gamma$   & 0.3889 & 0.066  &   \\
                            & $BB$ & 1.26 MeV & 20.39 & \\
                            & $BB^*$ & 4.58 MeV & 74.12 & \\
                            & $B^*B^*$ & 0.33 MeV & 5.34 & \\
                             &Total& 6.12 MeV & 100 & \\
\hline
\end{tabular}
\caption{Partial widths of radiative transitions and strong decays for 4D bottomonium mesons.}\label{results-10}
\end{table}

\begin{table}[h!]
\centering
\begin{tabular}{c c c c c c}
\hline
\hline
Initial State&Final State&Predicted Width(KeV)&Predicted B.R($\%$)\\
\hline\hline
$h_{b3} (1 ^1F_3) $         &  $\eta_{b2}(1 ^1D_2)\gamma$   & 12.7842 &  $\sim 100$    \\
                            &  $\chi_{b2}(1 ^3F_2)\gamma$   & $7.1654 \times 10^{-9}$ &  $\sim 0$   \\
                            &Total&12.7842 &100   \\
\hline
$\chi_{b2} (1 ^3F_2) $      &  $\Upsilon_1(1 ^3D_1)\gamma$   & 11.2759 & 84.86     \\
                            &  $\Upsilon_2(1 ^3D_2)\gamma$   & 1.9575 &  14.73    \\
                            &  $\Upsilon_3(1 ^3D_3)\gamma$   & 0.0537 &  0.4   \\
                            &Total&13.287 &100   \\
\hline
$\chi_{b3} (1 ^3F_3) $      &  $\Upsilon_2(1 ^3D_2)\gamma$   & 11.4229 & 89.28    \\
                            &  $\Upsilon_3(1 ^3D_3)\gamma$   & 1.3718 &  10.72    \\
                            &Total& 12.7947 &100   \\
\hline
$\chi_{b4} (1 ^3F_4) $      &  $\Upsilon_3(1 ^3D_3)\gamma$   & 12.3727 &  $\sim 100$    \\
                            &  $h_{b3}(1 ^1F_3)\gamma$   & $5.8055 \times 10^{-9}$ &  $\sim 0$   \\
                            &Total& 12.3727 &100   \\
\hline
\end{tabular}
\caption{Partial widths of radiative transitions and strong decays for 1F bottomonium mesons.}\label{results-11}
\end{table}

\section{Results and discussion}
\label{results and discussion}

We use the non-relativistic quark potential model to calculate the numerical wave functions and masses of bottomonium mesons. The mass spectrum of bottomonium mesons are calculated upto 2F energy states. A comparison of our predicted spectrum with recent theoretical studies and experimental data is reported in Table\ref{bottomonium masses}.

Our theoretical masses of bottomonium states Table\ref{bottomonium masses} show that 1S, 2S, 3S and 4S lying below the $BB$ threshold ($\approx10.558$GeV). Our theoretical mass of $4^3S_1$ is $10.437GeV$ lying below threshold but its experimental mass is $10.5794\pm0.0012GeV$ which is very close to $BB$ threshold. Our predicted width of $4^3S_1$ is $20.645MeV$ which is in good agreement with experimental width $20.5\pm2.5MeV$. The $\eta_b(5^1S_0)$ is not an established state and its predict mass is $10.6069GeV$ which is above threshold. According to spin selection rules and energy conservation $\eta_b(5^1S_0)$ has four open-bottom decay channels: $BB^*$, $B^*B^*$, $B_sB_s^*$ and $B_s^*B_s^*$. The predicted width of $\eta_b(5^1S_0)$  is $52.894MeV$. The $\Upsilon(5^3S_1)$ has six open-bottom decay channels: $BB$, $BB^*$, $B^*B^*$, $B_sB_s$ $B_sB_s^*$ and $B_s^*B_s^*$ with predicted width $50.47MeV$ which is in agreement with experimental width $37\pm4MeV$.

The 1P and 2P bottomonium states are experimentally established but lying below $BB$ threshold, therefore only radiative widths are calculated. The experimental masses of two multiplets of 3P bottomonium states are available whereas the masses of other two are not available experimentally. Our theoretical masses of $4^3P_2$ is very close to the $BB$ and has very small width of $0.01MeV$ which is not included in the tables.

The theoretical masses of 1D, 2D and 3D bottomonium states show that these states are below the $BB$ threshold whereas 4D states are above threshold. The $4^1D_2$ state decays strongly through $BB^*$ decay mode only with total predicted width is $4.839MeV$. The $4^3D_1$ state has two open-bottom decay modes $BB$ and $BB^*$  with total predicted width is $3.2MeV$. The predicted width of $4^3D_2$ multiplet is $3.41MeV$ with $BB^*$ decay mode only.  The $4^3D_3$ bottomonium state can decay strongly through $BB$, $BB^*$ and $B^*B^*$ decay channels with total predicted width is $6.12MeV$.

We have also included the theoretical masses of 1F and 2F bottomonium states in Table\ref{bottomonium masses} even that the higher states of bottomonium states are not experimentally established. According to our theoretical predictions 1F and 2F states are lying below $BB$ threshold and can decay through E1 and M1 transitions only.

Our predicted widths in Tables(\ref{results-1}-\ref{results-11}) show that the M1 radiative widths are very small, but E1 radiative widths are higher values up to 21.75 keV.  The reason of this difference is that M1 radiative widths depend on the factor $(\frac{1}{m^2_b})$ while this factor is not used in the calculation of E1 radiative widths.  Tables (4-14) show that the branching ratios of radiative widths are high below threshold, while the branching ratios of radiative widths decrease above threshold because of the existence of strong decays. Similar behavior is observed in refs.\cite{Godfrey15,wang18}.


\end{document}